\documentstyle[12pt]{article}
\begin{document}

\begin{center}

             {\bf    AN UPPER BOUND ON THE HIGGS BOSON MASS FROM A POSITIVITY CONDITION ON THE MASS MATRIX}\\
\mbox{}\\

		     Ali Al - Naghmoush, Murat \"Ozer  and M. O. Taha\\

			Department of Physics, College of Science\\
				  King Saud University\\
                        P.O. Box 2455, Riyadh 11451, Saudi Arabia \\                    
\mbox{}\\
				{\bf  ABSTRACT}
\end{center}
	We impose the condition that the eigenvalues of the mass matrix in the shifted Lagrangian density be positive at $\phi=\phi_{0}$, the vacuum expectation value of the scalar field. Using the one - loop effective potential of the standard model, this condition leads to an upper bound  on the Higgs boson mass $m_{H} :m_{H} < 230 GeV$, for a top quark mass of $175 GeV$.

\newpage

     	Recent experimental and theoretical work has considerably restricted the range over which the Higgs boson mass, of the minimal standard electroweak model (SM), may be found. The best direct experimental evidence is of a lower limit $m_{H}>65GeV (95\% C.L.)$  from CERN [1]. This work is reviewed in ref.[2]. The theoretical lower limits are not as firm. They arise from imposing the requirement of vacuum stability in a perturbative context. The Linde - Weinberg bound [3] -$m_{H}>7GeV$  -has recently been considerably modified. The authors of references [4] and [5] obtain $m_{H}>116GeV$, while (for $m_{t}=175GeV$) the results of ref.[6] yield $m_{H}>152GeV$.

There is unfortunately no direct experimental upper bound on $m_{H}$. Indirect evidence for an upper bound arises from recent statistical analysis of precision electroweak data. References [7,8] give $m_{H}<309GeV$ on the basis of this data. Theoretically, there are really no firm upper bounds on  $m_{H}$ in SM, apart from the unitarity bound [9,10],$m_{H}\leq1TeV$; beyond the reach of the present generation of accelerators. The reason for the lack of firm theoretical upper bounds on $m_{H}$ is the apparent absence of convincing restrictive conditions on SM parameters. The requirement that SM couplings remain perturbative up to Planck scale has, for example, often been imposed to obtain upper bounds on $m_{H}$. The speculative hypothesis that a renormalizable theory is consistent only for zero coupling had also been used [11,12] to restrict $m_{H}$ from above. Such considerations can only indicate orders of magnitude and cannot yield reliable numbers.

In this letter we obtain an upper bound on $m_{H}$ by imposing a condition that has not been considered before and which, we believe, is very firm and generally acceptable. We require that the eigenvalues of the \mbox{mass - matrix} (which we call eigenmasses) in the shifted Lagrangian density be real at $\phi=\phi_{0}$, where $\phi=\phi_{0}$ is the absolute minimum of the effective potential (EP) . We consider the direct consequences of this criterion to be highly reliable. Any limitations on these consequences can only come from other extraneous postulates or approximations employed in addition to this criterion to obtain these consequences.

One can see that this condition is firm and reliable from the observation that, at least in the first few orders, the loop expansion of the EP in the SM, massive \mbox{$\phi^{4}$ - theory} and scalar electrodynamics, indicates that the EP depends logarithmically on the eigenmasses. It is this logarithmic dependence on the eigenmasses that induces an imaginary part in the EP over certain regions in $\phi$. We anticipate that these features of the \mbox{finite - order} loop expansion are in fact valid for the exact EP. Then the condition which we impose would necessarily imply that the EP is real at its absolute minimum. This is obviously an essential property of the EP if its absolute minimum characterizes the vacuum state of the theory [13]. The value of $V(\phi_{0})$ is the vacuum energy of the system.

For the EP of the minimal SM the squared eigenmasses are 
\begin{displaymath}
M_{1}^{2}=-m^{2}+\frac{1}{2}\lambda\phi^{2}\hspace{0.5cm},\hspace{0.5cm} 
M_{2}^{2}=-m^{2}+\frac{1}{6}\lambda\phi^{2}\hspace{0.5cm},\hspace{0.5cm}
M_{3}^{2}=\frac{1}{4}g^{2}\phi^{2}\hspace{0.2cm}, 
\end{displaymath}
\begin{equation} 
M_{4}^{2}=\frac{1}{4}(g^{2}+g'^{2})\phi^{2}\hspace{0.5cm},\hspace{0.5cm}
M_{5}^{2}=\frac{1}{2}h^{2}\phi^{2}\hspace{0.2cm},
\end{equation}
where $g$ and $g'$ are the gauge couplings and $h$ is the top quark Yukawa coupling. Other fermion couplings are neglected. The mass parameter $m$ and the scalar coupling $\lambda$ occur in the potential 
\begin{equation} 
V_{0}(\phi)=-\frac{1}{2}m^{2}\phi^{2}+\frac{\lambda}{4!}\phi^{4}
\hspace{0.2cm},
\end{equation}
with $m^{2}>0$.

Due to the extensive data taken at the $Z^{0}$ resonance, the physical values of the gauge couplings are given at $\mu=M_{Z}$. We shall therefore take $\mu=M_{Z}$ in the expression that we use for the EP $V(\phi)$. The parameters $m^{2}$ and $\lambda$ would then also assume their values at $\mu=M_{Z}$. One is then left with the coupling $\lambda$ as the only unknown parameter in $V(\phi)$, since the values of the top quark mass $m_{t}$ and the vacuum expectation value $\phi_{0}$ of the Higgs field are known and the condition $\left(\partial V/\partial \phi \right)_{\phi=\phi_{0}}=0$ determines $m^{2}$ as a function of $\lambda$. The parameter $\lambda$, evaluated at $\mu=M_{Z}$, is not the physical coupling. One may, however, express both $m_{H}^{2}(=(\partial^2V/\partial\phi^{2})_{\phi=\phi_{0}})$    and $\lambda_{phys.}(=(\partial^4V/\partial\phi^4)_{\phi=\phi_{0}})$ in terms of $\lambda(M_{Z})$. Any restriction on $\lambda(M_{Z})$  is therefore a restriction on these physical values. The fact that this restriction is evaluated at $\mu=M_{Z}$ is immaterial for the upper bound on $m_{H}$.
 
The conditions $M_{k}^{2}(\phi=\phi_{0})>0 (k=1,2,...5)$ are satisfied for all $k$ when
\begin{equation} 
\lambda\phi_{0}^{2}>6m^{2} \hspace{0.2cm}.
\end{equation}
This is the exact necessary condition that must be satisfied so that all eigenmasses are real at the physical point $\phi=\phi_{0}$. It follows from the Lagrangian density and guarantees that the EP is real at its absolute minimum. Since $m^{2}$ is a function of $\lambda$ and the other known couplings, this condition is a restriction on the possible values of $\lambda$.

Condition (3) is exact. However, in determining $m^{2}$ as a function of the couplings, we need to use approximate schemes. In this letter we use the \mbox{one - loop} EP $V(\phi)=V_{0}(\phi)+V_{1}(\phi)$ of the SM in the \mbox{'t Hooft - Landau} gauge and the $\overline{MS}$ scheme, where $V_{0}$ is given in Eq.(2) and [6,14]
\begin{equation} 
V_{1}=\kappa \sum_{k=1}^{5}a_{k}M_{k}^{4}(ln\frac{M_{k}^{2}}{\mu^{2}}+b_{k})
\hspace{0.2cm},
\end{equation}
where
\begin{displaymath}
\kappa=(16\pi^{2})^{-1}\hspace{0.2cm},\hspace{0.2cm}a_{1}=\frac{1}{4}
\hspace{0.2cm},\hspace{0.2cm}a_{2}=\frac{3}{4}\hspace{0.2cm},\hspace{0.2cm}
a_{3}=-3\hspace{0.2cm},\hspace{0.2cm}a_{4}=\frac{3}{2}\hspace{0.2cm},\hspace{0.5cm}a_{5}=\frac{3}{4}\hspace{0.2cm},
\end{displaymath}
\begin{equation} 
b_{1}=b_{2}=b_{3}=-\frac{3}{2}\hspace{0.2cm},\hspace{0.2cm}b_{4}=b_{5}=-\frac{5}{6}\hspace{0.2cm}.
\end{equation}
The condition $(\partial V/\partial\phi)_{\phi=\phi_{0}}=0$ is then 
\begin{equation} 
-m^{2}+\frac{1}{6}\lambda\phi_{0}^{2}+4\kappa\sum_{k=1}^{5}a_{k}d_{k}M_{k}^{2}(\phi_{0})[ln\frac{M_{k}^{2}(\phi_{0})}{\mu^{2}}+b_{k}+\frac{1}{2}]=0
\hspace{0.2cm},
\end{equation}
where we have written $M_{k}^{2}(\phi)=c_{k}m^{2}+d_{k}\phi^{2}$. Define the function $f$ of $m^{2}/\mu^{2}$ and the couplings by
\begin{equation} 
f=\frac{1}{6}\frac{\lambda\phi_{0}^{2}}{m^{2}}\hspace{0.2cm}.
\end{equation}
In terms of this function Eq.(6) becomes
\begin{equation} 
f=1-4\kappa\sum_{k=1}^{5}a_{k}d_{k}\left(c_{k}+\frac{6d_{k}}{\lambda}
f\right)\left[ln\left(c_{k}+\frac{6d_{k}}{\lambda}f\right)+b_{k}+
\frac{1}{2}+ln\left(\frac{m^2}{\mu^2}\right)\right]\hspace{0.2cm},
\end{equation} 
and condition (3) is simply $f>1$. Thus the boundary $f=1$ of the region in which condition (3) is satisfied is given by
\begin{equation} 
\sum_{k\neq 2}a_{k}d_{k}\left(c_{k}+\frac{6d_{k}}{\lambda}
f\right)\left[ln\left(c_{k}+\frac{6d_{k}}{\lambda}f\right)+b_{k}+
\frac{1}{2}+ln\left(\frac{\lambda\phi_{0}^{2}}{6M_{Z}^{2}}\right)\right]=0
\hspace{0.2cm},
\end{equation} 	           
where we have set $\mu=M_{Z}$. This condition determines the limiting value  $\lambda=\lambda_{L}$ of the Higgs coupling for the region in which the reality condition (3) is satisfied. Whether this constitutes an upper bound or a lower bound on $\lambda_{0}$, the value of $\lambda$ at $\mu=M_{Z}$, is determined by the behavior of $f$ in the neighborhood of  $\lambda=\lambda_{L}$, noting that $f=1$ at $\lambda=\lambda_{L}$. With the following numerical values at $\mu=M_{Z}$ [6] 
\begin{displaymath}
g_{0}=0.650 \hspace{0.2cm},\hspace{0.2cm}g'_{0}=0.358 \hspace{0.2cm},\hspace{0.2cm} \phi_{0}=246GeV \hspace{0.2cm},
\end{displaymath}
\begin{equation} 
M_{Z}=91GeV \hspace{0.2cm},\hspace{0.2cm}h_{0}=\sqrt{2}\frac{m_{t}}{\phi_{0}} \hspace{0.2cm},
\end{equation} 					
where $m_{t}$ is the top - quark mass, we find $\lambda_{L}=2.643\pm 0.284$  for $m_{t}=175\pm 6GeV$ [15]. The variation of $f$ in the neighborhood of    $\lambda=\lambda_{L}$ shows that $f$ decreases for $\lambda>\lambda_{L}$    so that $\lambda_{L}$ is an upper bound on $\lambda_{0}$. We have also checked the numerical value of the ratio $\left |V_{1}(\phi=\phi_{0},\lambda=\lambda_{L})/V_{0}(\phi=\phi_{0},\lambda=\lambda_{L})\right |$  and found it $\approx 0.01$.
This indicates that the \mbox{one - loop} expression is a reliable approximation to the EP for these values of the parameters. A posteriori, justification for  the validity of the perturbative  expansion also follows from noting that $\lambda_{L}/4\pi\ll1$.

With this value of $\lambda_{L}$  one may now calculate the value of  the Higgs boson mass $m_{H}$, from $m_{H}^{2}=(\partial^{2}V/\partial\phi^{2})_{\phi=\phi_{0}}$, using the one - loop EP. This calculation establishes our result:
\begin{equation} 
m_{H}<230_{-13}^{+12}GeV\hspace{0.2cm},
\end{equation} 						   		
for $m_{t}=175\pm 6GeV$. As a  check, we also calculated the upper bound on the basis of the  value of $m_{H}$  obtained from the \mbox{tree - level} expression $m_{H}^{2}=\lambda_{0}\phi_{0}^{2}/3$. The upper bound is in this case  larger than that in (11) by less than $0.5\%$. Considering the results of recent analysis of the precision electroweak data [7,8], one concludes that our upper bound is quite relevant to the present experimental situation. When the result of ref.[6] is taken together with ours, one restricts the possible range for  to the rather narrow \mbox{experimentally - accessible} interval
\begin{equation} 
150GeV<m_{H}<240GeV
\end{equation}

Obvious generalizations of our work are the investigation of higher loop effects and the use of modified or improved forms of the \mbox{one - loop} expression for the EP. One may also attempt to use nonperturbative representations of the EP [16]. 

 {\bf References }

1. ALEPH Collaboration, D. Buskulic et al., Phys. Lett. {\bf B313} (1993) 299;
DELPHI Collaboration, P. Abreu et al., Nucl. Phys. {\bf B421} (1994) 3 ;
OPAL Collaboration, M. Z. Akrawy et al., Phys. Lett. {\bf B327} (1994) 397.\\ 
2. A. Sopczak, CERN report CERN - PPE/95 - 46 (1995).\\
3. A. D. Linde, JETP Lett. {\bf 23} (1976) 64;
 S. Weinberg, Phys. Rev. Lett. {\bf 36} (1976) 294.\\
4. G. Altarelli and I. Isidori, Phys. Lett. {\bf B337} (1994) 141.\\
5. J. A. Casas, J. R. Espinosa and M. Quiros, Phys. Lett. {\bf B342} (1995) 171;
J. R. Espinosa and M. Quiros, Phys. Lett. {\bf B353} (1995) 257.\\
6. C. Ford, D. R. T. Jones, P. W. Stephenson and M. B. Einhorn, Nucl. Phys. {\bf B 395} (1993) 17.\\
7. A. Blondel, report at the International Conference on High Energy Physics, Warsaw, 1996.\\
8. J. Ellis, G. L. Fogli and E. Lisi, CERN report CERN - TH/96-216.\\
9. C. Vayonakis, Lett. Nuov. Cim. {\bf 17} (1976) 383.\\
10. B. W. Lee, C. Quigg and H. B. Thacker, PR D16 (1977) 1519.\\
11. L. Maiani, G. Parisi and R. Petronzio, Nucl. Phys. {\bf B136} (1978) 115.\\
12. M. Lindner, Z. Phys. {\bf C31} (1986) 295.\\
13. S. Coleman and E. Weinberg, Phys. Rev. {\bf D7} (1973) 1888.\\
14 C. Ford, I. Jack and D. R. T. Jones, Nucl. Phys. {\bf B387} (1992) 373.\\
15. D. Gerdes, report hep-ex/9609013.\\
16. See, for example, P. M. Stevenson, Phys. Rev. {\bf D30} (1984) 1712.

\end{document}